\documentclass{jpsj2}

\usepackage{bm}
\usepackage{graphics}
\usepackage{graphicx}
\usepackage{eepic}
\newcommand{\e}{{\rm e}}
\renewcommand{\d}{{\rm d}}
\renewcommand{\i}{{\rm i}}

\title{Superconductivity in the vicinity of charge ordered state in
organic conductor $\beta$-({\it meso}-DMBEDT-TTF)$_2$PF$_6$}

\author{Kazuyoshi \textsc{Yoshimi}$^1$
\thanks{E-mail address: yoshimi@issp.u-tokyo.ac.jp},
Masaaki \textsc{Nakamura}$^2$ and Hatsumi \textsc{Mori}$^1$}

\inst{$^1$Institute for Solid State Physics, University of Tokyo,
 Kashiwanoha, Kashiwa, Chiba 277-8581\\
 $^2$Department of Applied Physics, Faculty of Science,
 Tokyo University of Science,
 Kagurazaka, Shinjuku-ku, Tokyo 162-8601}

\abst{
 We study theoretically competition between the charge ordering and the
 superconductivity in the two-dimensional organic conductor
 $\beta$-({\it meso}-DMBEDT-TTF)$_2$PF$_6$. We analyze the extended Hubbard
 model on a weakly-dimerized lattice based on the random phase
 approximation and Eliashberg equations. We found the reentrant behavior of
 the checkerbord-type charge-ordered phase in the phase diagram, and the
 triplet superconductivity due to the charge fluctuation in the
 neighboring region.
 In low temperatures, the singlet superconductivity also appears due to the
 enhancement of the spin fluctuation.
}

\kword
{
 organic conductor,
 $\beta$-({\it meso}-DMBEDT-TTF)$_2$PF$_6$,
 superconductivity,
 charge ordering,
 random phase approximation,
 Eliashberg equation,
 charge fluctuation
}

\def\runauthor{
K. \textsc{Yoshimi}, M. {\sc Nakamura} and
H. \textsc{Mori}}

\begin{document}
\maketitle

\section{Introduction}

Organic conductors have attracted attention for many years.  About two
decades ago, superconductivity was found in the low dimensional organic
conductor (TMTSF)$_{2}$PF$_6$ at the critical temperature
$T_{\rm{c}}\approx$ 0.9 K under the pressure $P\approx 12$ kbar. Since
then, about 130 kinds of organic superconductors have been discovered.
It is notable that BEDT-TTF salts (ET salts) are about 50 kinds of
them\cite{Ishiguro,Seo-H-F}. The ET salts are mainly
quasi-two-dimensional (2D) conductors with $3/4$-filled $\pi$ band of
the donor molecules.  These materials exhibit very interesting
electronic properties such as superconductivity, magnetism and charge
ordering.

For example, $\kappa$-(ET)$_2$X salts show superconducting (SC)
transitions next to the antiferromagnetic state in the $(P,T)$ phase
diagram. The theoretical calculations based on the half-filled extended
Hubbard model where the strong dimerization is assumed show that this
superconductivity is mediated by antiferromagnetic spin
fluctuation\cite{H.Kino,H.Kondo,M.Vojita,T.Jujo}. In
$\alpha$-(ET)$_2$I$_3$ salt, there are several theoretical works for the
superconductivity in the presence of a charge ordering by the random
phase approximation (RPA)\cite{A.Kobayashi1,A.Kobayashi2}.  In this case, 
the SC state is also considered to be mediated by the antiferromagnetic spin
fluctuation. For $\theta$-(ET)$_2$X salts where the various
charge-ordered (CO) states are observed, there are several theoretical
works on the CO states by the mean field theory\cite{H.Seo,M.Kaneko}.
The superconductivity of this material is explained that it is induced
by both the spin and the charge fluctuations\cite{Y.Tanaka}.

\begin{figure}[t]
\begin{center}
     \resizebox{80mm}{!}{\includegraphics{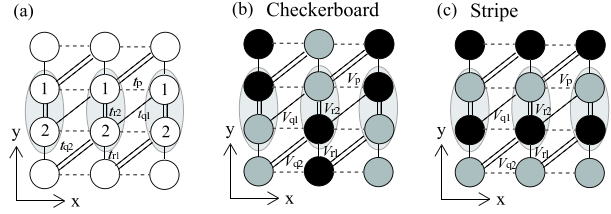}}
  \end{center}
 \caption{(a) Model of the conduction layer for
 $\beta$-(DMeET)$_2$PF$_6$ salt. The unit cell contains two molecules
 labeled by 1,2 with transfer integrals: $t_{r1}$, $t_{r2}$, $t_{q1}$,
 $t_{q2}$ and $t_{p}$. We also introduce corresponding nearest neighbor
 Coulomb interactions: $V_{r1}$, $V_{r2}$, $V_{q1}$, $V_{q2}$ and $V_{p}$. Two candidates 
 of the charge ordered states in the present analysis, (b) the checkerboard and (c) the stripe type CO states, respectively. The black and gray molecules are in charge-rich and poor states.}
 \label{fig:transfer_order.eps}
\end{figure}

In this paper, we focus on $\beta$-({\it meso}-DMBEDT-TTF)$_2$PF$_6$
salt ($\beta$-(DMeET)$_2$PF$_6$ salt) which has been recently
synthesized and investigated by Kimura and Mori {\it et al.}
\cite{S.Kimura,S.Kimura2,H.Mori} The conduction layer of this material
consists of weakly dimerized two molecules per a unit cell with the
$3/4$-filled band. This salt shows the metal-insulator (MI) transition
at $T_{\rm MI} \approx 90$ K without the anomaly of the magnetic
susceptibility around $T_{\rm MI}$.  The insulating state exhibits the
checkerboard-type charge ordering (see
Fig.{\ref{fig:transfer_order.eps}}(b)) where the charge disproportion
occurs within the dimer unit. Estimating the nearest neighbor Coulomb
interaction by the point charge approximation, it is reported that the
checkerboard type charge ordering cannot be explained, so that the
effective molecular size and the electron-phonon interaction should be
taken into account.\cite{S.Kimura} The most remarkable point is that the
SC state of $\beta$-(DMeET)$_2$PF$_6$ is next to the CO state in the
$(P,T)$ phase diagram, with the critical temperature $T_{\rm{c}}
\approx$ 4 K under the pressure $P\approx 4$ kbar \cite{S.Kimura2,
H.Mori}.  $T_{\rm{c}}$ decreases with increasing the magnetic field, and
the superconducting transition is completely suppressed above 4 T down
to 0.5 K.

It is well known that SC transitions often appear next to the magnetic
ordered region of the $(P,T)$ phase diagram not only in organics but
also in heavy fermion systems.\cite{Jaccard,Mathur,Saxena} This has been
considered as a typical character of superconductivity with magnetically
mediated pairing mechanism. Therefore, the fact that 
superconductivity appeared in the neighbor of the CO region is not only
rare for organic conductors, but also important for possibilities of
the superconductivity mediated by the charge fluctuation. Recently, SC state 
appeared in the neighboring region of the CO state 
is also found in $\beta$-vanadium bronze\cite{T.Yamauchi}, but
no theoretical studies have been carried out.

From the above background, we study $\beta$-(DMeET)$_2$PF$_6$ in the
following two point of views: i) how the checkerboard-type charge
ordering is stabilized in the weakly dimerized organic system. ii) how
the SC region appears in the neighbor of the CO region from the metallic state.

The rest of this paper is organized as follows. In Sec.~2, we formulate
susceptibilities and Eliashberg equations based on the RPA for the
weakly dimerized 2D system. In Sec.~3, we determine appropriate
parameter set of the model, and present results of the numerical
analysis. Finally, summary and discussion are given in Sec.~4.

\section{Formulation}
In this section, we perform the analysis for the Hubbard-type model
based on the random phase approximation following Kobayashi {\it et
al.}\cite{A.Kobayashi1,A.Kobayashi2}. Throughout this paper we set
$k_{\rm B}=\hbar=1$.

\subsection{Model}
\begin{figure}[t]
\begin{center}
     \resizebox{80mm}{!}{\includegraphics{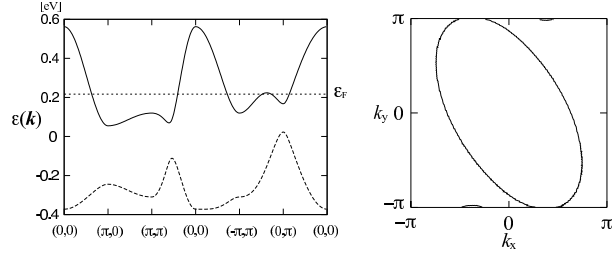}}
  \end{center}
 \caption{Energy band spectra $\xi_{1}$ and $\xi_{2}$ of eq.~(\ref{E_HF}) and the Fermi surface.}
 \label{fig:band.eps}
\end{figure}

We apply the extended Hubbard model on the two-dimensional (2D)
lattice (see Fig.~\ref{fig:transfer_order.eps} (a)),
\begin{align}
 {\cal H}&={\cal H}_t+{\cal H}_U+{\cal H}_V, \label{Hubbard1}\\
 {\cal H}_{t}&=\sum_{\langle i\alpha;j\beta \rangle}\sum_{\sigma}
  (t_{i\alpha;j\beta }c_{i\alpha\sigma}^{\dag}
  c_{j\beta\sigma}^{\mathstrut}+\mbox{H.c.}),\\
 {\cal H}_{U}&=\sum_{i\alpha}U_{\alpha}
  n_{i\alpha\uparrow}n_{i\alpha\downarrow},\\
 {\cal H}_{V}&=\sum_{\langle i\alpha;j\beta\rangle}\sum_{\sigma\sigma'}
  V_{i\alpha;j\beta}n_{i\alpha\sigma}n_{j\beta\sigma'},
\end{align}
where $i,j$ ($\in 1,\cdots, N_{L}$) denote the lattice points of the 2D
square lattice, and $\alpha,\beta$ ($\in 1,2$) specify molecules in the
unit cell. $c_{i\alpha\sigma}^{\dag}$ ($c_{i\alpha\sigma}$) is the
creation (annihilation) operator for an electron with spin $\sigma$
($=\uparrow,\downarrow$). $\langle i\alpha;j\beta \rangle$ represents a
bond pair between the nearest neighbor sites. $t_{i\alpha;j\beta}$
denotes the transfer energy between sites $(i,\alpha)$ and $(j,
\beta)$. ${\cal H}_{U}$ and ${\cal H}_{V}$ denote repulsive interactions
where $U_{\alpha}$ and $V_{\alpha\beta}$ are the coupling for the
on-site and those for the nearest neighbor sites, respectively.  By
using the Fourier transformation,
\begin{equation}
 c_{i\alpha\sigma}
  =\frac{1}{\sqrt{N_L}}\sum_{\bm{k}}
  \e^{\i \bm{k}\cdot \bm{r}_{i}}c_{\bm{k},\alpha,\sigma}^{\mathstrut},
\end{equation}
eq.~($\ref{Hubbard1}$) is rewritten as
\begin{align}
 \lefteqn{{\cal H}=\sum_{\bm{k}\alpha\beta\sigma}
 (\varepsilon_{\alpha\beta}(\bm{k})c_{\bm{k}\alpha\sigma}^{\dag}
 c_{\bm{k}\beta\sigma}^{\mathstrut}+\mbox{H.c.})} \nonumber\\
&+\frac{1}{N_L}\sum_{\bm{k}\bm{k}'\bm{q}\alpha}
 U_{\alpha}c_{\bm{k}+\bm{q},\alpha,\uparrow}^{\dag}
 c_{\bm{k},\alpha,\uparrow}^{\mathstrut}
 c_{\bm{k}'-\bm{q},\alpha,\downarrow}^{\dag}
 c_{\bm{k}',\alpha,\downarrow}^{\mathstrut}\nonumber\\
&+\frac{1}{2N_L}\sum_{\bm{k}\bm{k}'\bm{q}\alpha\beta\sigma\sigma'}
 V_{\alpha\beta}(\bm{q})
 c_{\bm{k}+\bm{q},\alpha,\sigma}^{\dag}
  c_{\bm{k},\alpha,\sigma}^{\mathstrut}
  c_{\bm{k}'-\bm{q},\beta,\sigma'}^{\dag}
  c_{\bm{k}',\beta,\sigma'}^{\mathstrut}
 \label{Hubbard2}  
\end{align}
where the parameters in eq.~(\ref{Hubbard2}) are taken as, $\varepsilon
_{11}=\varepsilon_{22}$,
$\varepsilon_{12}^{\mathstrut}=\varepsilon_{21}^{*}$, $V_{11}=V_{22}$,
$V_{12}^{\mathstrut}=V_{21}^{*}$. Then the transfer energies are
\begin{align}
 \varepsilon_{11}(\bm{k})&=2t_{p}\cos k_{x},\\
 \varepsilon_{12}(\bm{k})&=t_{r1}\e^{-\i k_{y}}+t_{r2}
  +t_{q1}\e^{\i k_x}+t_{q2}\e^{-\i(k_{x}+k_{y})}.
\end{align}
The long range Coulomb interactions become
\begin{align}
 V_{11}(\bm{q})&=2V_{p}\cos q_{x},\\
 V_{12}(\bm{q})&=V_{r1}\e^{-\i q_{y}}+V_{r2}
  +V_{q1}\e^{\i q_x}+V_{q2}\e^{-\i(q_{x}+q_{y})}.
\end{align}
Applying the mean field approximation, the Hamiltonian is given by
\begin{align}
 {\cal H}_{\rm{MF}}&=\sum_{\bm{k}\alpha\beta\sigma}
  \tilde{\varepsilon}_{\alpha\beta\sigma}(\bm{k})
  c_{\bm{k}\alpha\sigma}^{\dag}
  c_{\bm{k}\beta\sigma}^{\mathstrut}+\mbox{const}.,\label{H_MF}\\
 \tilde{\varepsilon}_{\alpha\beta\sigma}(\bm{k})
  &=\varepsilon_{\alpha\beta}(\bm{k})
  +\delta_{\alpha\beta}
  \left[
  U_{\alpha}\langle n_{\alpha\bar{\sigma}}\rangle 
   +\sum_{\beta'\sigma'}V_{\alpha\beta'}
   \langle n_{\beta'\sigma'}\rangle\right],\nonumber\\
\end{align}
where $\bar{\sigma}$ denotes the opposite spin of $\sigma$. In this
approximation, we ignore the Fock term ($\langle
c_{i\alpha}^{\dag}c_{j\beta}^{\mathstrut}\rangle \simeq 0$) and assume
the uniform ground state: $\langle n_{\alpha \sigma}\rangle=3/4$.  The
Hamiltonian is diagonalized as
\begin{equation}
 {E}_{\rm{MF}}=\sum_{\bm{k}\alpha\gamma\sigma}\xi_{\gamma\sigma}(\bm{k})
  d_{\alpha\gamma\sigma}^{*}(\bm{k})
  d_{\alpha\gamma\sigma}^{\mathstrut}(\bm{k}),\label{E_HF}
\end{equation}
where $\xi_{1}(\bm{k})>\xi_{2}(\bm{k})$ ( see Fig.~\ref{fig:band.eps}
). $d_{\alpha\gamma\sigma}(\bm{k})$ is the element of unitary matrix
obtained by diagonalization of eq.~(\ref{H_MF}). Then, $\langle
n_{\alpha\sigma} \rangle$ is given as
\begin{equation}
 \langle n_{\alpha\sigma} \rangle=\sum_{\gamma=1}^{2}
  \frac{
  d_{\alpha\gamma\sigma}^{*}(\bm{k})
  d_{\alpha\gamma\sigma}^{\mathstrut}(\bm{k})
  }{\exp\left[(\xi_{\gamma\sigma}(\bm{k})-\mu)/T\right]+1},
\end{equation}
where $\mu$ is the chemical potential determined by the condition of
$3/4$-filling: $ \frac{1}{2}\sum_{\alpha\sigma}\langle n_{\alpha
\sigma}\rangle =\frac{3}{2}$.

\subsection{Charge and spin susceptibilities}

We define matrices of susceptibility as
\begin{eqnarray}
 \lefteqn{
 \left(X_{\sigma\sigma'}(\bm{q},\i\omega_{n})\right)_{\alpha\beta}\equiv}
 \nonumber\\
&&\frac{1}{N_{L}}\int_{0}^{1/T}\d\tau\,
 \e^{\i\omega_{n}\tau}
 \langle n_{\alpha\sigma}(\bm{q})(\tau)
 n_{\beta\sigma'}(-\bm{q})(0) \rangle,
 \label{X_def}
\end{eqnarray}
where $\omega_l$ is the Matsubara frequency for
bosons\cite{A.Kobayashi1}. $\tau$ is the imaginary time.  The density
operator $n_{\alpha,\sigma}(\bm{q})$ is given by
\begin{equation}
 n_{\alpha,\sigma}(\bm{q})=\sum_{\bm{k}}
  c_{\bm{k},\alpha,\sigma}^{\dag}
  c_{\bm{k}+\bm{q},\alpha,\sigma}^{\mathstrut}.
\end{equation}
The charge susceptibility $\hat{X}^{\rm c}$ and the spin susceptibility
 for the easy axis $\hat{X}^{\rm s}$ are given by
\begin{align}
\hat{X}^{\rm c}_{\alpha\beta} &= \frac{1}{2}(\hat{X}_{\uparrow\uparrow}
 +\hat{X}_{\downarrow\uparrow}+\hat{X}_{\uparrow\downarrow}
 +\hat{X}_{\downarrow\downarrow})_{\alpha\beta},\\
\hat{X}^{\rm s}_{\alpha\beta} &= \frac{1}{2}(\hat{X}_{\uparrow\uparrow}
 -\hat{X}_{\downarrow\uparrow}-\hat{X}_{\uparrow\downarrow}
 +\hat{X}_{\downarrow\downarrow})_{\alpha\beta}.
\end{align}

Applying the random phase approximation and ignoring the Fock term, we
 obtain
\begin{align}
\hat{X}_{\sigma\sigma}&=
\big[ I+\hat{X}^{(0)}\hat{V}^{(1)}-\hat{X}^{(0)}
\hat{V}^{(2)}(I+\hat{X}^{(0)}\hat{V}^{(1)})^{-1}\nonumber\\
&\times\hat{X}^{(0)}\hat{V}^{(2)}\big] ^{-1}\hat{X}^{(0)},\\
\hat{X}_{\bar{\sigma}\sigma}
&=-(I+\hat{X}^{(0)}\hat{V}^{(1)})^{-1}\hat{X}^{(0)}\hat{V}^{(2)}
\big[I+\hat{X}^{(0)}\hat{V}^{(1)}\nonumber\\
&-\hat{X}^{(0)}\hat{V}^{(2)}(I+\hat{X}^{(0)}\hat{V}^{(1)})^{-1}
\hat{X}^{(0)}\hat{V}^{(2)}\big]^{-1}\hat{X}^{(0)}.
\end{align}
$\hat{V}^{(1)}$ and $\hat{V}^{(2)}$ are given by
\begin{equation}
\hat{V}^{(1)}=\hat{V}+\hat{U},\qquad
\hat{V}^{(2)}=\hat{V}
\end{equation}
where $\hat{V}$ and $\hat{U}$ are defined as
\begin{equation}
\hat{V}=
\left(
  \begin{array}{cc}
    V_{11}(\bm{q})   &  V_{12}(\bm{q})\\
    V_{21}(\bm{q})   &  V_{22}(\bm{q})
  \end{array}
\right),\qquad
\hat{U}=
\left(
  \begin{array}{cc}
    U   &  0  \\
    0   &  U
  \end{array}
\right).
\end{equation}
In this case, we treat mainly the nonmagnetic state, i.e.,
$X_{\uparrow\uparrow}=X_{\downarrow\downarrow}$,
$X_{\uparrow\downarrow}=X_{\downarrow\uparrow}$.  Then $\hat{X}^{\rm c}$
and $\hat{X}^{\rm s}$ are given by
\begin{align}
 \hat{X}^{\rm c}
  &=\hat{X}_{\uparrow\uparrow}+\hat{X}_{\uparrow\downarrow}\nonumber\\
 &=(\hat{I}+\hat{X}^{(0)}(\hat{U}+2\hat{V}))^{-1}\hat{X}^{(0)},
  \label{Xc_RPA}\\
 \hat{X}^{\rm s}
  &=\hat{X}_{\uparrow\uparrow}-\hat{X}_{\uparrow\downarrow}\nonumber\\
 &=(\hat{I}-\hat{X}^{(0)}\hat{U})^{-1}\hat{X}^{(0)},
  \label{Xs_RPA}
\end{align}
where $\hat{X}^{\rm c}$ and $\hat{X}^{\rm s}$ are hermitian matrices.
The irreducible susceptibility $\hat{X}^{(0)}$ is
\begin{align*}
 X_{\alpha\beta}^{(0)}(\bm{q},\i\omega_n)=
  -\frac{T}{N_L}
  \sum_{\bm{k},n}
  G_{\alpha\beta}^{(0)}(\bm{k}+\bm{q},\i\omega_m+\i\epsilon_{n})
  G_{\beta\alpha}^{(0)}(\bm{k},\i\epsilon_{n})
\end{align*}
where $\epsilon_{n}$ denotes the Matsubara frequency for
fermions. $G_{\alpha\beta}^{(0)}(\bm{k},\i\omega_{n})$ is the single
particle Green function given by
\begin{equation}
 G_{\alpha\beta}^{(0)}(\bm{k},\i\epsilon_{n})=
  \sum_{\gamma=1}^2
  \frac{d_{\alpha\gamma}^{\mathstrut}(\bm{k})d_{\beta\gamma}^{*}(\bm{k})
  }{\i\epsilon_{n}-\xi_{\gamma}(\bm{k})+\mu},
  \label{Green_func}
\end{equation}
where $\xi_{\gamma}(\bm{k})$ and $d_{\alpha,\gamma}(\bm{k})$ are defined
in eq.~($\ref{E_HF}$).

Now, we introduce the linear combinations of the density operators as
\begin{align}
 n_{\pm}(\bm{q})=\frac{1}{\sqrt{2}}
  \left(n_{1}(\bm{q})\pm n_{2}(\bm{q})\right).
\end{align}
Then replacing the original density operators in eq.~(\ref{X_def}), and
taking account the symmetry in the unit cell, the corresponding charge
($\nu=$ c) and spin ($\nu=$ s) susceptibilities are given by
\begin{align}
 \hat{X}^{\nu}_{\pm}&=
  \frac{1}{2}(\hat{X}^{\nu}_{11}\pm\hat{X}^{\nu}_{21}
  \pm\hat{X}^{\nu}_{12}+\hat{X}^{\nu}_{22})\nonumber\\
 &=\hat{X}^{\nu}_{11}\pm {\rm Re\,}\hat{X}^{\nu}_{12}.
\end{align}
The stability of the CO states in the present analysis (see
Fig.{\ref{fig:transfer_order.eps}} (b), (c)) is given by divergence of
$\hat{X}_{-}^{\rm c}$.

\subsection{Pairing interactions and Eliashberg equations}
The superconducting transition point is determined by the Eliashberg
equation. The linearized Eliashberg equation for the singlet SC state is
given by
\begin{eqnarray}
 &&\lambda_{\rm S}\Sigma_{\alpha\sigma;\beta\bar{\sigma}}^{\rm a}(\bm{k})
  =-\frac{1}{N_{L}}\sum_{\bm{k}',n',\alpha',\beta'}
  P_{\alpha\sigma;\beta\bar{\sigma}}^{\rm S}(\bm{k}-\bm{k}')\nonumber\\
 &&\times
 G_{\alpha\alpha'}^{(0)}(\bm{k}',\i\epsilon_{n'})
 G_{\beta\beta'}^{(0)}(-\bm{k}',-\i\epsilon_{n'})
  \Sigma_{\alpha'\sigma;\beta'\bar{\sigma}}^{\rm a}(\bm{k}'),
\label{EL_sing1}
\end{eqnarray}
with the paring interaction,
\begin{equation}
\hat{P}^{\rm S}=\hat{U}+\hat{V}
 +\frac{3}{2}\hat{U}\hat{X}^{\rm s}\hat{U}
 -\frac{1}{2}(\hat{U}+2\hat{V})\hat{X}^{\rm c}(\hat{U}+2\hat{V}),
\end{equation}
where $\hat{X}^{\rm c}$ and $\hat{X}^{\rm s}$ are the susceptibility
matrices given in eqs.(\ref{Xc_RPA}) and (\ref{Xs_RPA}).  In
eq.~(\ref{EL_sing1}), $\Sigma_{\alpha\sigma;\beta\bar{\sigma}}^{\rm
a}(\bm{k})$ is the anomalous self-energy with the space inversion symmetry
$\Sigma_{\alpha\sigma;\beta\bar{\sigma}}^{\rm a}(\bm{k})=
\Sigma_{\beta\sigma;\alpha\bar{\sigma}}^{\rm a}(-\bm{k})$.  Here we
neglect the dependence of the Matsubara frequency in
$\Sigma_{\alpha\sigma;\beta\bar{\sigma}}^{\rm a}(\bm{k})$, since such a
treatment is valid for the weak coupling case.\cite{Y.Tanaka} In
eq.~(\ref{EL_sing1}), $\lambda_{\rm S}=1$ corresponds to the
superconducting transition point.

Similarly, the linearized Eliashberg equation for the triplet SC state
is obtained as
\begin{eqnarray}
 &&\lambda_{\rm T}\Sigma_{\alpha\sigma;\beta\sigma}^{\rm a}(\bm{k})
  =-\frac{1}{N_{L}}\sum_{\bm{k}',n',\alpha',\beta'}
  P_{\alpha\sigma;\beta\sigma}^{\rm T}(\bm{k}-\bm{k}')\nonumber\\
 &&\times G_{\alpha\alpha'}^{(0)}(\bm{k}',\i\epsilon_{n'})
 G_{\beta\beta'}^{(0)}(-\bm{k}',-\i\epsilon_{n'})
 \Sigma_{\alpha'\sigma;\beta'\sigma}^{\rm a}(\bm{k}'),
 \label{EL_triplet}
\end{eqnarray}
with the paring interaction,
\begin{equation}
 \hat{P}^{\rm T}=\hat{V}-\frac{1}{2}\hat{U}\hat{X}^{\rm s}\hat{U}
  -\frac{1}{2}(\hat{U}+2\hat{V})\hat{X}^{\rm c}(\hat{U}+2\hat{V}).
  \label{PT_eq}
\end{equation}

In order to confirm the relevant part of the effective interaction for
the singlet state, we divide the paring interaction into the following
two parts,
\begin{align}
\hat{P}^{\rm c}&=\hat{V}-\frac{1}{2}(\hat{U}+2\hat{V})
 \hat{X}^{\rm c}(\hat{U}+2\hat{V}),\label{Pc}\\
\hat{P}^{\rm s}&=\hat{U}+\frac{3}{2}\hat{U}\hat{X}^{\rm s}\hat{U}.
 \label{Ps}
\end{align}

In the present linearized Eliashberg equations, the amplitude of the
anomalous energies can be chosen arbitrary, so that we calculate the
quasi-particle bands just below the SC transition temperature.  The
quasi-particle bands is obtained by the following Hamiltonian,
\begin{align}
 &{\cal H}_{\rm{MF}}^{\rm SC}
  ={\cal H}_{\rm{MF}}-\mu
  \sum_{\bm{k}}\sum_{\sigma}\sum_{\alpha}
  c_{\bm{k}\alpha\sigma}^{\dag}c_{\bm{k}\alpha\sigma}^{\mathstrut}
 \nonumber\\
 &+\lambda(
  \Sigma_{\alpha\beta}^{\rm a}(\bm{k})
  c_{\bm{k}\alpha\sigma}^{\dag}c_{-\bm{k}\beta \sigma _1}^{\dag}
  +\Sigma_{\alpha\beta}^{\rm a \ *}(\bm{k})
  c_{-\bm{k}\beta \sigma _1}c_{\bm{k}\alpha\sigma} )\bigr].
 \label{gap_eq}
\end{align}
where $\sigma_{1}=\bar{\sigma}$ $(\sigma_{1}=\sigma)$ for the singlet
(triplet) SC state.

\section{Numerical Results}
Since there are a lot of parameters in the present model, we select a
set of parameters in the following way. First, we fix the transfer
integrals calculated by the extended H\"{u}ckel method based upon the
crystal structure analysis: $t_{r1}=-0.0824, t_{r2}=-0.226,
t_{p}=0.0475, t_{q1}=-0.0438$ and $t_{q2}=-0.115 \
[\mbox{eV}]$.\cite{S.Kimura2} In this case, we do not consider explicit
pressure dependence of the parameters. Next, we determine the value of
the on-site Coulomb energy $U$ which becomes the ground state
nonmagnetic.  Then we determine the value of the nearest neighbor
interactions $V$ which stabilize the checkerboard type CO state.  We use
the unit for the energies as [eV].  Finally, we determine the phase
diagram of this model.

\subsection{Nonmagnetic state}
\begin{figure}[t]
\begin{center}
     \resizebox{80mm}{!}{\includegraphics{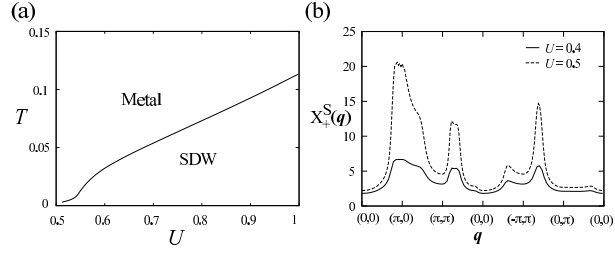}}
  \end{center}
  \caption{(a) Phase diagram in the $(U,T)$ plane. The solid line
 correspond to the phase boundary between the metallic and the SDW
 states. (b) Spin susceptibilities $X^{\rm s}_{+}(\bm{q})$ for $U=0.4$
 and $0.5$ at $T=0.01$.}  \label{fig:UT.eps}
\end{figure}

To obtain the phase diagram in the $(U,T)$ plane
(Fig.~\ref{fig:UT.eps}(a)), we calculate $X^{\rm s}_{+}$ which is larger
than $X^{\rm s}_{-}$.  Figure~\ref{fig:UT.eps}(b) shows the momentum
dependence of the spin susceptibilities $X^{\rm s}_{+}$, at
$T=0.1$. When the system is dimerized, the exchange coupling
$J=\frac{4t^2}{U}$ becomes $J_{x}>J_{xy}>J_{y}$, where
$|t_{x}|=|t_{p}-t_{q1}/2|=0.069 > |t_{xy}|=|t_{q2}|/2=0.056 >
|t_{y}|=|t_{r1}/2|=0.0412$.\cite{Seo-H-F} As a result, $X^{\rm s}_{+}$
becomes the largest at ($\pi, 0$). $\beta$-$(\mbox{DMeET})_2\mbox{PF}_6$
 is nonmagnetic. Therefore, we fix the on-site Coulomb interaction $U$ at 
 $0.4$ [eV]  in the following calculation where the spin
density wave (SDW) does not appear.


\subsection{Charge ordered states}

\begin{figure}[t]
\begin{center}
     \resizebox{80mm}{!}{\includegraphics{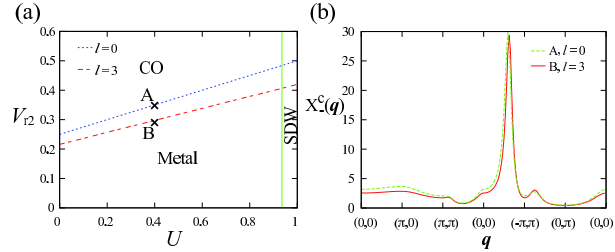}}
  \end{center}
  \caption{(a) Phase diagram on $(U,V_{r2})$ plane at $T=0.1$. The dashed
    and the solid lines correspond to the CO and the SDW
    instabilities, respectively. (b) Charge susceptibilities $X^{\rm
    c}_{-}(\bm{q})$ for $(V_{r2},l)=(0.35,0),(0.294,3)$ at $T=0.1,
    U=0.4$.}\label{fig:phase_UV.eps}
\end{figure}

We investigate the combination of the nearest neighbor Coulomb
interactions $V$, where the checkerboard type charge order is
stabilized.  First, we examine the charge order patterns by using the
point charge approximation: $V_{\i}\propto
\frac{1}{r_{\i}}$. Figure~\ref{fig:phase_UV.eps}(a) shows the phase
diagram on the ($U$,\ $V_{r2}$) plane at $T=0.1$. $\hat{X}^{\rm c}_{-}$
becomes larger in the intermediate region between ($0,0$) and
($-\pi,\pi$) as shown in Fig.~\ref{fig:phase_UV.eps}(b), therefore we
find that this result does not accord with the experimental fact. Taking
account of the effective molecular size $l$, the coulomb interaction in
the dimer is given as $V_{r2}\propto (r_{r2}^2+l^2)^{-1/2}$. It is
reported that $l\approx 3$ \AA\ is reasonable for the ET
salt.\cite{M.Tamura} Therefore we fix $l$ at $3$ \AA\ and apply the point
charge approximation for the other $V$s. Figure~\ref{fig:phase_UV.eps}(a)
shows ($U$,\ $V_{r2}$)-phase diagram at $T =0.1$.  In the case of
$l=3$ \AA, the CO instability occurs in the
smaller $V_{r2}$ than that of $l=0$ \AA. This indicates that the charge
fluctuation becomes larger by taking account of the molecular size,
because the difference between $V_{r2}$ and another $V$ is
smaller. However, the critical point of $X^{\rm c}_{-}(\bm{q})$ does not
change and we cannot obtain the checkerboard type charge ordering.

In order to stabilize the checkerboard type CO state, it is suitable to
satisfy the following condition:
$(V_{r2},V_{p},V_{q2})>(V_{r1},V_{q1})$. For simplicity, we assume only
two sets of nearest neighbor interactions $V_{1}(=V_{p}=V_{q2})$ and
$V_{2}(=V_{r1}=V_{q1})$. As shown in
Fig.~\ref{fig:transfer_order.eps}(b), $V_{1}$ is the interaction to induce
the charge disproportion with the charge density in the unit cell:
$n_{i1}\neq n_{i2}$. $V_{2}$ is the interaction to favor the uniform
charge density in the unit cell: $n_{i1}=n_{i2}$. We set
$V_{r2}=s_{1}\times U,\ V_{1}=V_{p}=V_{q2}=s_{2}\times V_{r2}, \
V_{2}=V_{r1}=V_{q1}=s_{3}\times V_{r2}$. For nonmagnetic states, we fix
$U = $ 0.4 eV and $V_{r2}= 0.8\times U$ as shown in
Fig.~\ref{fig:UT.eps}(a).

\begin{figure}[t]
 \begin{center}
  \begin{tabular}{cc}
   \resizebox{55mm}{!}{\includegraphics{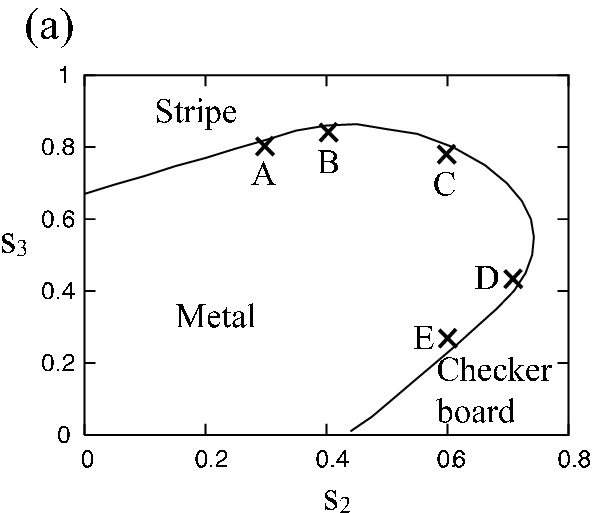}} \\
   \resizebox{50mm}{!}{\includegraphics{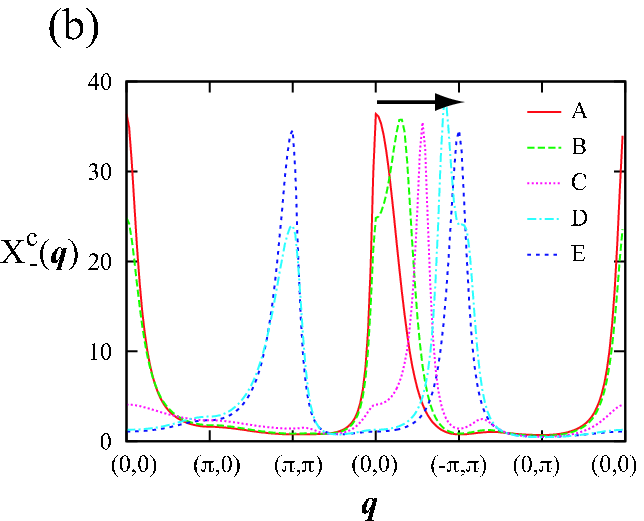}} \\
  \end{tabular}
  \caption{(a) Phase diagram on $(s_{2},s_{3})$ plane at $T=0.1, U=0.4,
  V_{r2}=0.8U$. The solid line corresponds to the CO instability. (b)
  Charge susceptibilities $X^{\rm c}_{-}(\bm{q})$ for
  $(s_{2},s_{3})=\rm{A}(0.3,0.8), \rm{B}(0.4,0.84), \rm{C}(0.6,0.78),
  \rm{D}(0.7,0.41)$ and $\rm{E}(0.6,0.25)$ at $T=0.1, U=0.4,
  V_{r2}=0.8U$. The arrow shows the shift of the peak of $X^{\rm
  c}_{-}(\bm{q})$ from $(0,0)$ to
  $(\pi,\pi)$}\label{fig:phase_s1s2_com.eps}
 \end{center}
\end{figure}

Figure~\ref{fig:phase_s1s2_com.eps} shows the phase diagram on the
$(s_{2},s_{3})$ plane. As shown in Fig.~\ref{fig:phase_s1s2_com.eps}(b), a peak of
$X^{\rm c}_{-}$ appears at $(0,0)$ in the stripe type CO phase
(Fig.~\ref{fig:transfer_order.eps}(c)), and $(\pi,\pi)$ in the
checkerboard type CO phase (Fig.~\ref{fig:transfer_order.eps}(b)).  The
wave number for the peak of $X^{\rm c}_{-}$ changes continuously in the
intermediate region between the stripe and the checkerboard type CO
regimes. In the parameter region of the point charge approximation,
$s_{2}$ is nearly equal to $s_{3}$, but $s_{2}$ should be sufficiently
larger than $s_{3}$ to realize the checkerboard type CO state.  When we
take account of the correspondence with the real material, it is
suitable that the difference between $s_{2}$ and $s_{3}$ is
small. Therefore, we set $s_{3}=0.3$.

\begin{figure}[tbp]
\begin{center}
     \resizebox{80mm}{!}{\includegraphics{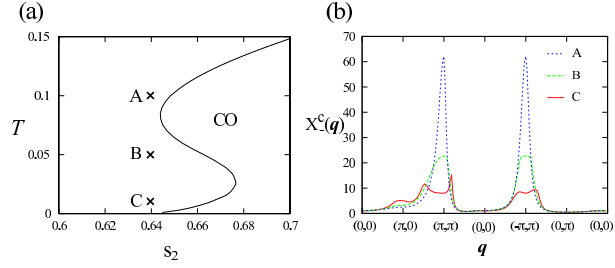}}
  \end{center}
  \caption{(a) Phase diagram for the CO state on $(s_{2},T)$ plane at
  $U=0.4, V_{r2}=0.8U,V_{2}=0.3V_{r2}$. The solid line corresponds to
  the CO instability. (b) Charge susceptibilities $X^{\rm
  c}_{-}(\bm{q})$ for $T=$ (A) $0.1$, (B) $0.05$ and (C) $0.01$ at $U=0.4,
  V_{r2}=0.8U,V_{1}=0.64V_{r2}$ and $V_{2}=0.3V_{r2}$.}
    \label{fig:phase_CDW.eps}
\end{figure}

Next, we examine the CO state by varying the temperature. The obtained
phase diagram is shown in Fig.~\ref{fig:phase_CDW.eps}(a). There is a
reentrant charge order transition as a function of temperature similar
to the result of the extended Hubbard model in the 2D square
lattice\cite{A.Kobayashi3}.  We consider that this reentrant behavior is
due to the competition between the effect of $V(\bm{q})$ and
$X^{(0)}(\bm{q})$. At high temperatures, the peak of $X^{\rm c}_{-}$ is
near $(\pi,\pi)$. This originates from the momentum dependence of
$V(\bm{q})$. With decreasing temperatures, the momentum dependence of
$X^{(0)}(\bm{q})$ becomes large.  Thus the peak of $X^{\rm
c}_{-}(\bm{q})$ changes from $(\pi,\pi)$ as shown in
Fig.~\ref{fig:phase_CDW.eps} (b).

\subsection{Pairing interactions and superconducting states}
\begin{figure}[t]
  \begin{center}
    \begin{tabular}{cc}
      \resizebox{50mm}{!}{\includegraphics{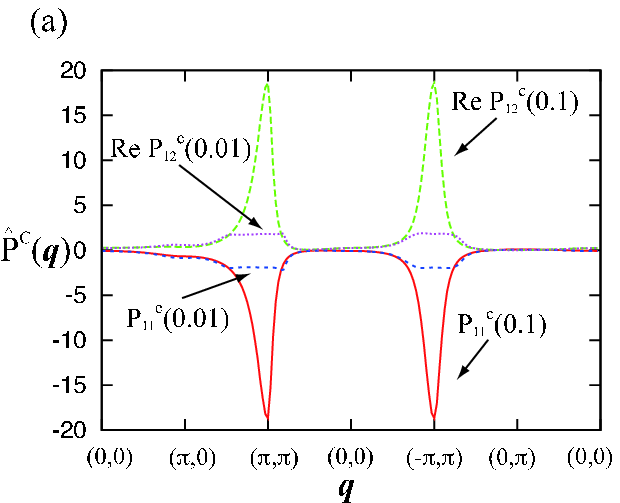}} \\
      \resizebox{50mm}{!}{\includegraphics{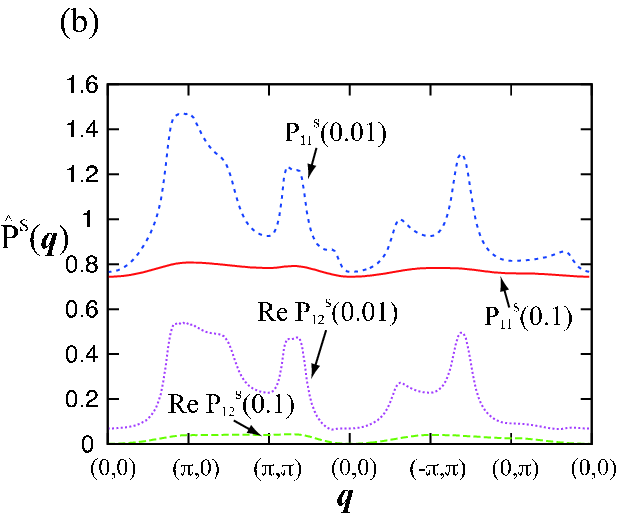}} \\
    \end{tabular}
    \caption{Momentum dependence of the pairing interactions for the (a)
   charge [eq.~(\ref{Pc})] and the (b) spin parts [eq.~(\ref{Ps})],
   $P^{\rm c(s)}_{\alpha\beta}(T)$, at $U=0.4,V_{r2}=0.8U$ and
   $V_{2}=0.3V_{r2}$. We set $s_2=0.64$ at $T$ = 0.1 and $s_2=0.613$ at
   $T$ = 0.01, respectively.}\label{fig:Pair_Tdep.eps}
     \end{center}
\end{figure}

First, we examine the effect of $\hat{P}^{\rm
c}$. Figure~\ref{fig:Pair_Tdep.eps}(a) shows the momentum dependence of
$\hat{P}^{\rm c}$. At $(s_{2},T)=(0.64,0.1)$, $\hat{P}^{\rm c}_{11}$
shows a negative peak at $(\pi,\pi)$, whereas $\hat{P}^{\rm c}_{12}$
shows a positive peak. This result is consistent with the appearance of
the checkerboard-type charge ordering. The effective repulsive
interaction is caused by the Coulomb interactions where $V_{r2}$ and
$V_{1}$ are larger than $V_2$.  On the other hand, the effective
attractive interaction arises by other Coulomb interactions $V_{2}$. At
$(s_{2},T)=(0.64,0.01)$, the peak of $\hat{P}^{\rm c}$ is suppressed and
changes from $(\pi, \pi)$. We think it originates from the reentrant
behavior of the CO.

The momentum dependence of $P^{\rm s}$ is determined by that of
$\hat{X}^{(0)}$, because $U$ does not depend on the momentum. As shown
in Fig.~\ref{fig:Pair_Tdep.eps}(b), the momentum dependence of $P^{\rm
s}$ increases with lowering temperatures. It is notable that
$\hat{P}^{\rm c}$ is much larger than $\hat{P}^{\rm s}$ at $T=0.1$
meaning that the SC state originates form the charge
fluctuation. However, at $T=0.01$, both $\hat{P}^{\rm c}$ and
$\hat{P^{\rm s}}$ are the same order. Therefore we expect that the SC
state in this region is caused by both the charge and the spin
fluctuations.
\begin{figure}[tbp]
\begin{center}
     \resizebox{80mm}{!}{\includegraphics{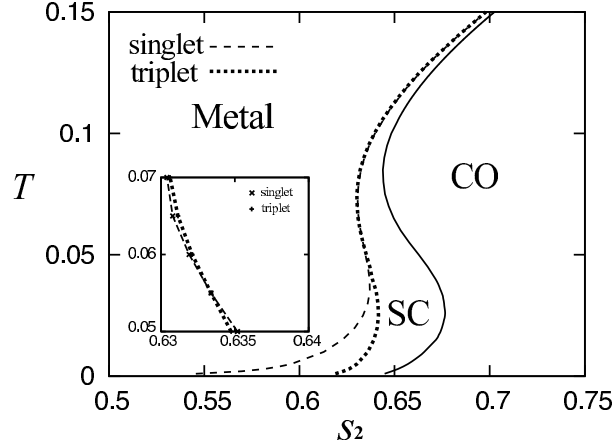}}
  \end{center}
  \caption{Phase diagram on $(s_{2},T)$ plane at
  $U=0.4, V_{r2}=0.8U, V_{2}=0.3V_{r2}$. The solid line corresponds to
  the CO instability. Note that the superconducting (SC) region is
  overestimated in the present analysis for the high-temperature
  region.The inset shows the detail close to $T = 0.06$ where the
  triplet SC state competes with the singlet SC state.}
  \label{fig:phase_SC.eps}
\end{figure}

\begin{figure}[tbp]
\begin{center}
     \resizebox{80mm}{!}{\includegraphics{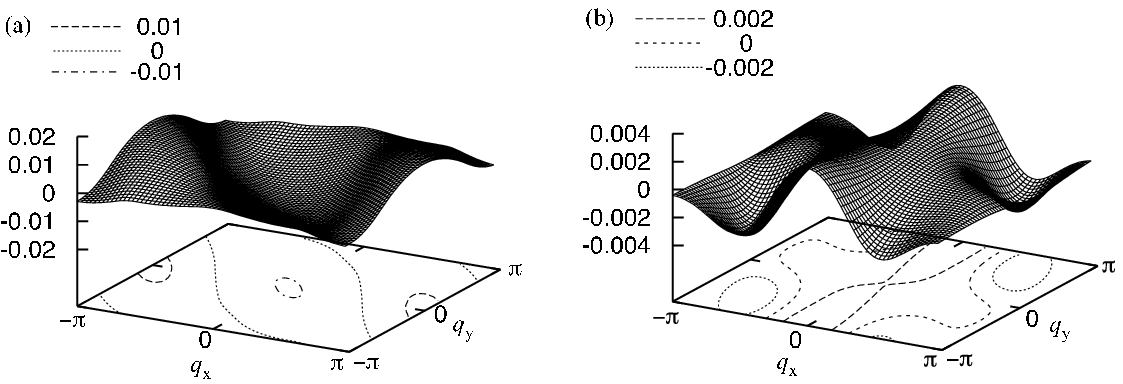}}
  \end{center}
  \caption{The momentum dependence of the element of anomalous
  self-energy matrix $\Sigma_{11}$ of the singlet SC state for (a)
  $T$=0.01, $s_{2}$=0.613 and (b) $T$=0.1, $s_2$=0.64. }
  \label{PS_delta_11.eps}
\end{figure}
\begin{figure}[tbp]
\begin{center}
     \resizebox{80mm}{!}{\includegraphics{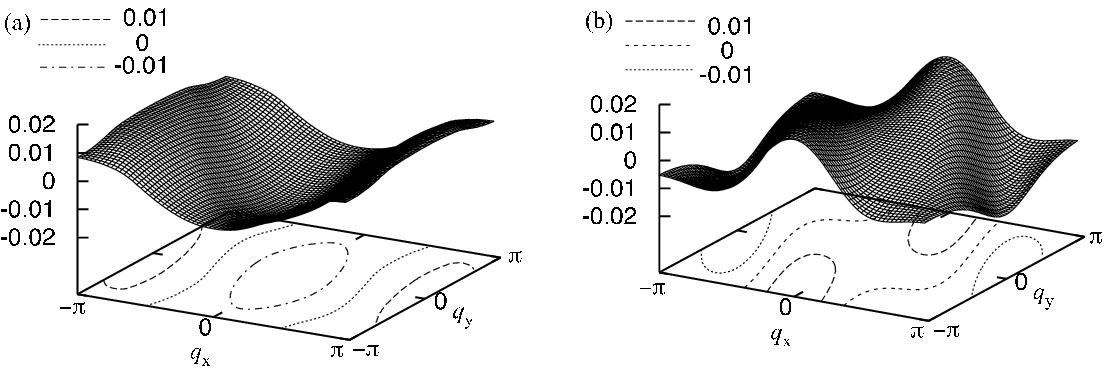}}
  \end{center}
  \caption{The momentum dependence of the element of anomalous
  self-energy matrix Re$\Sigma_{12}$ of the singlet SC state for (a)
  $T$=0.01, $s_{2}$=0.613 and (b) $T$=0.1, $s_2$=0.64.}
  \label{PS_Re_delta_12.eps}
\end{figure}
\begin{figure}[tbp]
\begin{center}
     \resizebox{80mm}{!}{\includegraphics{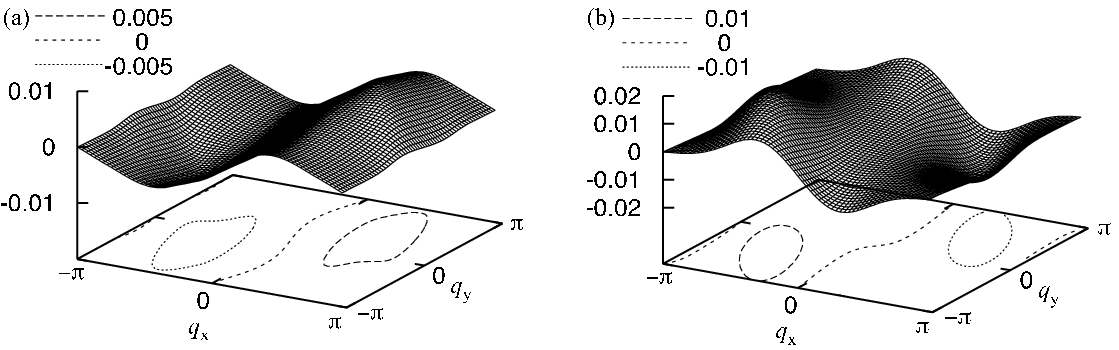}}
  \end{center}
  \caption{The momentum dependence of the element of anomalous
  self-energy matrix Im$\Sigma_{12}$ of the singlet SC state for (a)
  $T$=0.01, $s_{2}$=0.613 and (b) $T$=0.1, $s_2$=0.64.}
  \label{PS_Im_delta_12.eps}
\end{figure}

\begin{figure}[tbp]
\begin{center}
     \resizebox{80mm}{!}{\includegraphics{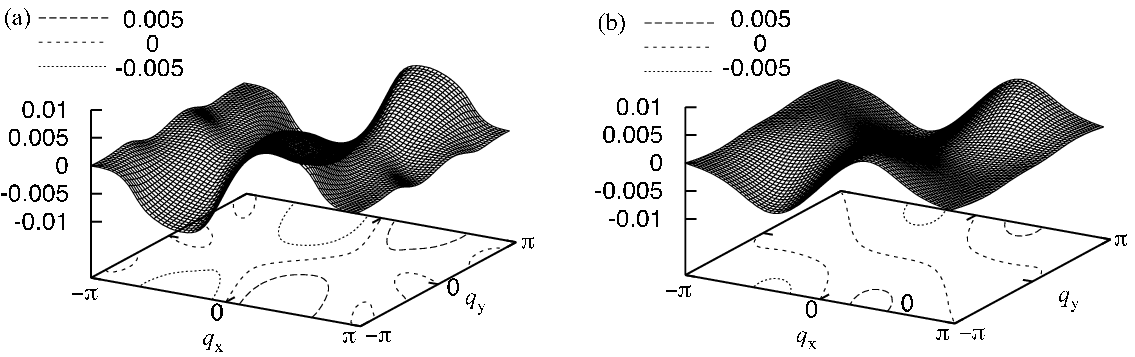}}
  \end{center}
  \caption{The momentum dependence of the element of anomalous
  self-energy matrix $\Sigma_{11}$ of the triplet SC state for (a)
  $T$=0.01, $s_{2}$=0.636 and (b) $T$=0.1, $s_2$=0.64.}
  \label{PT_delta_11.eps}
\end{figure}
\begin{figure}[tbp]
\begin{center}
     \resizebox{80mm}{!}{\includegraphics{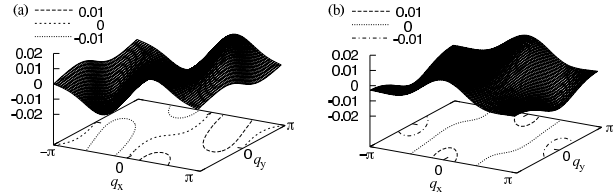}}
  \end{center}
  \caption{The momentum dependence of the element of anomalous
  self-energy matrix (a) Re $\Sigma_{12}$ 
  and (b) Im $\Sigma_{12}$ of the triplet SC state for $T$=0.1, $s_2$=0.64 .}
  \label{PT_delta_12.eps}
\end{figure}

In order to estimate the onset temperature for the SC states, we
evaluate the linearized Eliashberg equations eqs.~(\ref{EL_sing1}),
(\ref{EL_triplet}).  Then we obtain the phase diagram on the $(s_2,T)$
plane as shown in Fig.~\ref{fig:phase_SC.eps}, the singlet SC state
competes with the triplet SC state for $0.05<T$, but the singlet SC
transition occurs first for $T<0.05$.  This means that an incommensurate
CO phase is developed as lowering temperatures.

As shown in Figs.~\ref{PS_delta_11.eps} and \ref{PS_Re_delta_12.eps},
the anomalous self-energy of the singlet SC state strongly depends on
the temperature. For the singlet SC state at $T=0.1$, Re $\Sigma_{12}$
is larger than $\Sigma_{11}$. This indicates that Re $\Sigma_{12}$
contributes mainly to the SC state at $T=0.1$.  On the other hand, Re
$\Sigma_{12}$ is the same order of $\Sigma_{11}$.  This indicates that
both $\Sigma_{11}$ and Re $\Sigma_{12}$ contribute to the SC state at
$T=0.01$. As mentioned above, the spin fluctuation increases with
lowering temperatures (Fig.~\ref{fig:Pair_Tdep.eps}) and $P^{\rm s}$
is the same order of $P^{\rm c}$ at $T=0.01$. The pairing interactions
do not have a special peak, since the anomalous self-energies are broad.

The anomalous self-energy of the triplet SC does not depend much on the
temperature as shown in Fig.~\ref{PT_delta_11.eps}. Re $P^{\rm s}_{12}$
is smaller than $P^{\rm s}_{11} $. It follows from eq.~(\ref{PT_eq})
that the total pairing interaction of the triplet SC state is given by
$\hat{P}^{\rm T} = \hat{P}^{\rm c} - \frac{1}{3}(\hat{P}^{\rm s} -
\hat{U}) $. This indicates that the triplet SC state is less affected 
by the spin fluctuation than the singlet SC state. Therefore,
we consider that the triplet SC state is mainly induced by the charge
fluctuation. It is reasonable that the triplet SC state appears near the
CO state.

\begin{figure}[tbp]
\begin{center}
     \resizebox{60mm}{!}{\includegraphics{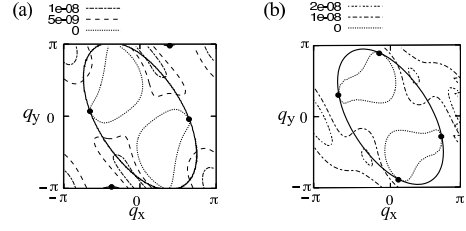}}
  \end{center}
  \caption{The gap of the singlet SC state from the quasi-particle band
  of the noninteracting case at (a) $T=0.01$ with $s_2=0.64$ and (b)
  $T=0.1$ with $s_2=0.613$. The solid and dashed line correspond to the
  Fermi line and the gap, respectively.}  \label{fig:Gap_singlet.eps}
\end{figure}

\begin{figure}[tbp]
\begin{center}
     \resizebox{40mm}{!}{\includegraphics{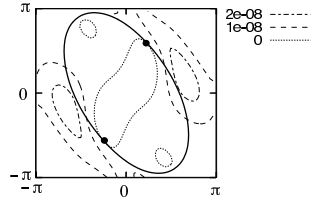}}
  \end{center}
  \caption{The gap of the triplet SC state from the quasi-particle band
  of the noninteracting case at $T=0.1$ with $s_2=0.613$. The solid and
  dashed line correspond to the Fermi line and the gap,
  respectively.}\label{fig:Gap_triplet.eps}
\end{figure}

 Finally, we determine the symmetry of the order parameters by
 calculating the quasi-particle bands and count the numbers of zero
 points of gaps at the Fermi surface.  We analyze eq.~(\ref{gap_eq}) as
 $\lambda \ll 1$: we set $\lambda =10^{-4}$ in this calculation, since
 the SC gap is regarded as infinitesimal. It is shown that the singlet
 SC state has 4 nodes as seen from Fig.~\ref{fig:Gap_singlet.eps}.  On
 the other hand, the triplet SC state has 2 nodes as shown in
 Fig.~\ref{fig:Gap_triplet.eps}.
 In this result, however, we found a pathological behavior of the
 triplet gap which does not vanish at the center of the momentum space
 $\bm{k}=(0,0)$, even though it satisfies the fermion antisymmetry.  In
 fact, in the present analysis, we cannot classify the singlet and
 triplet gaps as the parity in the momentum space. This reason is
 explained as follows:
 The symmetry of the singlet (triplet) state is given as $\Sigma^{\rm
 a}_{\alpha\beta}({\bm k})=\pm\Sigma^{\rm a}_{\beta\alpha}(-{\bm k})$,
 so that $\Sigma^{\rm a}_{\alpha\beta}({\bm k})=\pm\Sigma^{\rm
 a}_{\alpha\beta}(-{\bm k})$ is not satisfied in general for
 $\alpha\neq\beta$.  Therefore, the singlet (triplet) gap may include
 the parity-odd (-even) off-diagonal componets as shown in Fig.~\ref{PS_Im_delta_12.eps}
 (Fig.~\ref{PT_delta_12.eps}).
 In spite of this extra freedoms, the number of nodes coincides with
 the usual singlet- and triplet-gap properties. The explanation of this fact
 remains a future problem.

\section{Summary and Discussion}

We have theoretically examined the CO and the SC states of
$\beta$-(DMeET)$_2$PF$_6$. We have analyzed the extended Hubbard model
including the nearest Coulomb interactions in the weakly dimereized
lattice.  Applying the RPA, we choose the suitable parameters realizing
the checkerboard type CO state under the nonmagnetic state. The CO
phase shows reentrant behavior as a function of the temperature. We
consider that this behavior originates from the competition between the
wave vector dependence of $V$ and the nesting effect of the charge
susceptibility.

Using the pairing interactions induced by the charge and spin
fluctuations in terms of the RPA, we have estimated the onset
temperature of the SC state near the CO instability by the linearized
Eliashberg equations. The triplet SC state is stabilized around the
region of the CO instability which originates from the wave vector
dependence of $V(\bm{q})$.  On the other hand, the singlet SC state is
stabilized near the region of the CO instability which originates from
the nesting vector of the Fermi surface.

The determination of parameters of the present model has been carried out
 to realize the checkerboard type CO state under the nonmagnetic state,
comparing with the experimental result. However, values of these
parameters are much different from those expected by the point charge
approximation, even taking account of the effective molecular size.  This
suggests that the electron-phonon interaction due to the lattice distortion 
might be important to understand the checkerboard type CO state.


In this paper, we have assumed that the CO state is an insulating state.
There is a possibility, however, that the CO state is a metallic state
due to existence of the hole and the electron pocket in the Fermi
surface, and a SC state coexists with the CO state.  This scenario was
asserted to explain the superconductivity of $\alpha$-(ET)$_2$I$_3$ salt
under uniaxial pressure.\cite{A.Kobayashi1,A.Kobayashi2} Actually, the
temperature and the pressure dependences of the resistivity in certain
parameter regions seem to behave similarly in these two
systems.\cite{Tajima,S.Kimura2} If such a metallic state is realized, the
phase diagram of Fig.~\ref{fig:phase_SC.eps} may be modified so that a
different type of SC state appears in the CO region, but the SC states
obtained in the present analysis remain to be unchanged.

%

\section{Acknowledgements}
The authors thank Y. Suzumura and A. Kobayashi for discussion and many
helpful suggestions. We acknowledge T. Kato for helpful comments on the
mean field calculation, and Y. Tanaka for discussion about the numerical
analysis of the linearized Eliashberg equations.  We are also grateful
to H. Fukuyama and C. Ishii for useful comments.  One of the authors
(M.~N.)  is partly supported by the Grant-in-Aid for scientific research
of the Ministry of Education, Science, Sports and Culture of Japan.

\end{document}